# Towards a Formal, Visual Framework of Emergent Cognitive Development of Scholars

Amir Hussain and Muaz Niazi*

ahu@cs.stir.ac.uk , muaz.niazi@gmail.com

*Corresponding author

**Abstract** Understanding the cognitive evolution of researchers as they progress in the academia is an important but complex problem; a problem belonging to a class of problems, which often require the development of models for gaining further understanding in the intricacies of the domain. The research question that we address in this paper is how to effectively model this temporal cognitive mental development of prolific researchers. Our proposed solution to this problem is based on noting that the academic progression and notability of a researcher are linked with a progressive increase in the citation count for the scholar's refereed publications quantified using indices such as the Hirsch index. In other words, we propose the use of yearly cognitive increment of a scholar's cognition to be quantifiable using a function of the scholar's citation index, thereby considering the index as an indicator of the discrete approximation of the scholar's cognitive development. Using validated agent-based modeling, a paradigm presented as part of our previous work i.e. Cognitive Agent-based Computing framework, we present both formal as well as visual agent-based complex network representations for this cognitive evolution in the form of a Temporal Cognitive Level Network (TCLN) model. As a proof of the effectiveness of this approach, we demonstrate validation of the model using historic data of citations.

## 1. Introduction

In this era of proliferation of research publications and global availability of the internet, one way of noting the academic progression of a researcher in a discipline is by observing the progressive increase in the citation buildup of a scholar's refereed publications. What started out as an informal measure has now resulted in the formal usage of an increase in citations as a measure of notability of scholars (Weingart 2005). While some have argued about the exact nature of the significance of citations of literature (Amsterdamska and Leydesdorff 1989), if used intelligently, citations are known to serve as an indicator of the significance of published research (Cronin 1984). Perhaps a reason for the extensive usage of such metrics is that they allow for a discrete quantification of a person's research in terms of peer-evaluation.

While citations alone do not reflect a person's notability, citation buildup can eventually be captured by means of now well-known metrics such as the Hirsch index (Hirsch 2005) among others (Braun et al. 2006; Egghe 2010). Each of these indices attempts to give an approximate measure of the productivity, notability as well as impact of the published work of a scientist or scholar. While on one hand, the evolution of a scholar's Hirsch index represents the person's academic growth, on the other, it can perhaps also be considered as an indication of the cognitive mental evolution of the scholar. Thus, each yearly sampling of the h-index represents a discrete stage in the evolutionary cognitive development of the researcher.

**1.2 Contributions**

Could we somehow quantify this evolution in the form of a computational model? In general, this is the research question that we attempt to address in this paper. As a solution to this problem, we first develop a formal computational model for capturing this cognitive mental development of prolific researchers. Subsequently, using validated agent-based modeling level from the Cognitive Agent-based Computing framework (Niazi and Hussain 2012), we develop visual representations of this cognitive evolution followed by a proof of the effectiveness of this approach using historic data from citation indices.

Specifically, the contributions in this paper can be listed as follows:

1. We propose an approach to the modeling of cognitive development involved in scientific research by proposing a novel technique of merging citation data with researcher h-Index by proposing Temporal Cognitive Level Networks (TCLN), a network modeling approach allowing the visualization of evolutionary cognition of researchers by means of their research papers and associated Hirsch Index.
2. We also demonstrate the benefits of TCLN including a simpler visual representation as well as a significant reduction in the utilization of computational resources. In a TCLN representation, nodes representing researchers, their papers, their citations in addition to the emergent productivity of each researcher in the same representational model. It also requires a much lower order of memory footprint than existing author-paper network models.
3. As proof of concept, we present the design and development of an agent-based simulation model for TCLN. This model demonstrates both the calculation of Hirsch index as well as the representational capabilities of TCLN. Validation of the effectiveness of this approach is demonstrated by means of using historic data of researchers' evolution retrieved from the Google scholar Index. TCLN can thus represent researchers, papers, citations as well as research productivity in the same model. We also analyze the space complexity proving a reduction of complexity from the order of $O(mn)$ for author-paper networks to the linear order O(m), where "m" is the average number of papers published by a researcher and "n" is the number of researchers. The main objective of this work is to simplify the network representation such that it is easier to study the co-evolution of state and topology in research networks as well as represent researcher repute and productivity.

4. For a validation of the framework and Hirsch index calculations, we design and develop an agent-based model, which is then compared with a well-known tool "Publish or Perish" ([Harzing and van der Wai 2008](#)) to retrieve historical data from the Google scholar index. The agent-based simulation model also demonstrates the representational power of TCLNs.

**1.3 Outline**

The structure of the rest of the paper is as follows: In the next sub-section, first we provide background and related work. Next, we develop a formal framework and methodology using a formal specification language "Z" ([Hayden et al. 1968](#)). Subsequently, we describe the agent-based model and algorithms. This is followed by a discussion of the simulation experiments and validation exercises. Finally, we conclude the paper.

## 2. Background and related work

In this section, we present background and related work. We first start with the important.

### 2.1 Overview

An effect of the hosting of scientific communication (especially open access publications) over the web is the emergence of citation indices ([Falagas et al. 2008](#)) such as Thomson Reuters Web of Science, Scopus, Google Scholar and PubMed etc. Never before were we able to thereby examine a cross-sectional view of the research productivity of human civilization at any given time. While all such data is easily accessible online, the analysis of such data in terms of looking for specific patterns can be quite daunting because of the exponentially increasing size of citation corpora. As an example, a search for just the top 10 journals listed in "Computer Science, Cybernetics" category of Journal Citation Reports from the years 1995-2010 using the ISI web of knowledge gives result of the order of 2,801 papers, 6,376 authors, 4,787

keywords and 83, 633 cited references[1]. Whereas a topic search for "expert system*" returns 48,875 records.

Due to the complex nature of the citation data, simple statistical measures may not give ample useful information. Instead, an effective mechanism may be to use complex network approach, which involves the transformation of citation data into a network format. This complex network transformation is not optional since the particular network format is needed for the discovery of complex patterns. Now, there are several ways in which citation data can be represented in the form of a network. The simplest could be to create what is typically termed as a citation network (Hummon and Dereian 1989). In a citation network, entities such as research papers, authors, institutions or journals become nodes (or vertices[2]) and citations become the arcs (directed lines) or links. Another possibility is to develop a co-authorship network, which is based on making authors as nodes and co-authorships as lines (Newman 2004; Börner et al. 2004).

Whatever the means of development of the exact type of networks, the eventual goal is to be able to apply network analysis techniques on the resultant network. However, citation networks have certain representational and technical problems. Firstly, the size of citation corpora is growing exponentially. Representing it graphically for analysis would thus require an ever-increasing requirement of computational resources such as RAM. Typical methods using tools such as Network Workbench or Citespace (Chen 2006) solve this problem of network display by pruning the network to display top nodes only (Niazi and Hussain 2011). However, this type of pre-processing can actually possibly result in removal of papers/authors or other citation indicators which might have lesser citations on their own but might have served as central nodes in transition of ideas or connecting different subject categories. Secondly, traditional networks extracted from citation data do not represent the productivity of individual researchers because measuring productivity often requires the use of algorithms which need to take all publications into consideration and not just the top cited one.

### 2.1 Problems in Modeling Research

For development of a deeper understanding of the cognitive processes requires the development of explicit models (Epstein 2008) starting with implicit mental models. These models also serve the purpose of enhancing our knowledge about cognitive abilities while paving the way for future usage of ideas in the autonomous mental/cognitive development in robots and animals (Weng et al. 2001).

---

[1] Results based on search performed in August, 2010
[2] In literature, vertices and nodes are used interchangeably for graphs. Subsequently we shall only use the term "node".

Research can therefore be modeled as a Complex Adaptive System (CAS) (Niazi 2013) where numerous researchers at different levels of experience in performing research, work together to produce ideas valuable to the particular discipline. The evaluation of these ideas is conducted by means of the publication process of refereed research articles and patents. In addition, as time progresses, community interest, measurable by means of citations indices, indicates the strength of the proposed hypotheses in the views of other peer researchers. The research process (Niazi et al. 2008) is known to be tied in with numerous socio-cultural aspects as noted by Weng et al. (Weng et al. 2001). Watts and Gilbert (Watts and Gilbert 2011) have proposed a model of knowledge-seeking using through scientific publication. Fischer has proposed a theory of cognitive development, called the skill theory, where cognitive development is based on discrete skill sets.

As we can note here, all of these aspects primarily tie in with the ideas of complexity, evolution and in general, CAS. As such, while research has numerous aspects and facets which can be modeled at the micro levels, an intuitive way of modeling the emergent outcome of this process would be to examine the cognitive evolution of researchers, which is the goal of this paper. Here, the key problem addressed is how to develop a multi-faceted model comprising of computational, formal and visual representations of the cognitive mental development of researchers.

### 2.1  Cognitive Agent-based Computing

Recently agent-based modeling and complex networks have been combined and used in the form of Cognitive Agent-based computing, a single modeling and simulation framework presented in earlier in (Niazi 2011) and expanded in (Niazi and Hussain 2012). Applications of the framework have been shown in the areas of disaster alerting systems such as (Niazi et al. 2011), breast cancer decision support (Siddiqa et al. 2009), power saving and energy harvesting for corporate networks (Niazi and Laghari 2012), fault tolerance in large-scale transcational systems (Niazi et al. 2006) and peer-to-peer networks (Niazi 2008). Validated agent-based modeling using in the current paper has been presented earlier in (Niazi et al. 2009).

Other examples of agents and multiagent systems include the use of JADE agents such as by Fortino et al. in (Fortino et al. 2010). Likewise, an agent-based platform for programming Wireless sensor networks has been presented by Aiello et al. in (Aiello et al. 2011).

### 2.2  The importance of citations in research

"Publish or Perish" (Wikipedia 2008) refers to the pressure to publish work continuously to advance in the academia. However simply publishing articles is not enough. In general, the idea of research evaluation is based on assigning a scalar value directly or indirectly related to the citations of papers. These scalar values include the Thomson Reuters Journal Citation report Impact factors for Journals and various types of indices (Hirsch 2005; Braun et al. 2006; Egghe) proposed recently. Recent work (Egghe 2007) has shown that the Hirsch index is a concavely increasing function of time, asymptotically bounded by

$$T^{\frac{1}{\alpha}}$$

Where T is the total number of papers in that group and $\alpha$ is the exponent of Lotka's law of the system.

### 2.2.1 Impact of a paper

Citation analysis as a measure of productivity of researchers and evaluation of research has been studied by Moed (Moed 2005). The impact of a paper can be considered to be demonstrated by the interest shown in the paper by the community One way of measuring this is to count the number of non-self citations recorded in peer-reviewed scientific literature. The use of citations as a measure of impact can be attributed to Eugene Garfield, chairman emeritus ISI, who formalized and gave the idea of scientific impact and impact factors (Garfield 1955, 2006, 1972).

### 2.2.2 Impact of a researcher

The impact of a paper lead to the idea of measuring the impact of a scientist over the entire career. This measurement was first proposed in 2005 by Hirsch (Hirsch 2005). In general, the impact of a scientist can be measured by being highly cited by other authors. Egghe (Egghe 2010) presented a detailed overview of the inter-relation of Hirsch with related indices. These indices have been considered as effective measures of a scientist's impact or research productivity.

Hirsch index, which is the most fundamental of all indices is defined formally as:

"A scientist has index $h$ if $h$ of his/her $N_p$ papers have at least h citations each, and the other $(N_p - h)$ papers have no more than h citations each."

So, suppose an author has 6 papers: 5 of them have 4 citations and the sixth has 1 citation. Then $h = 4$. It is important to note here that the calculation of h-Index is not quite simple. It involves intelligent retrieval of information from some standard indexing source (such as Google Scholar or SCI, SSCI or Scopus etc.) and then calculation of the index. A well-known tool for the calculation of Hirsch index used extensively by researchers is Ann Harzing's "Publish or Perish" (Harzing and van der Wal 2008).

### 2.2.3 Traditional Citation Analysis Approach

Complex Networks are one way of representing relationships and interactions in CAS. In the domain of citation analysis, as noted by Hummon (Hummon and Dereian 1989), citation networks have been used to represent citations since Garfield's original paper (Garfield 1955). Research article networks are acyclic because all papers point back in time making close loops rare (Newman 2003; Batagelj 2003).

Nowadays, citation analysis (Chen and Redner 2010; Kajikawa and Takeda 2009) is performed using some of the following steps:
1. First, data is retrieved from citation databases such as Science Citation Index (SCI) and the Social Sciences Citation Index (SSCI) compiled by the Institute for Scientific Information (ISI). Data is filtered for scope and correctness.
2. Next, the data is pruned for top nodes
3. Next it is represented as a directed or undirected network.
4. Now, clusters are discovered from the data and highly cited and influential papers inside the cluster are identified using some algorithm (Clauset et al. 2004).
5. The network is drawn using various layouts.
6. Network is gradually modified to highlight particular structural features of interest.
7. Interrelation of citation clusters is then studied in detail.
8. Key nodes are identified using different types of centrality measures.

In the following Figure 1, we can see how a simple citation network can become with just a few thousand nodes.

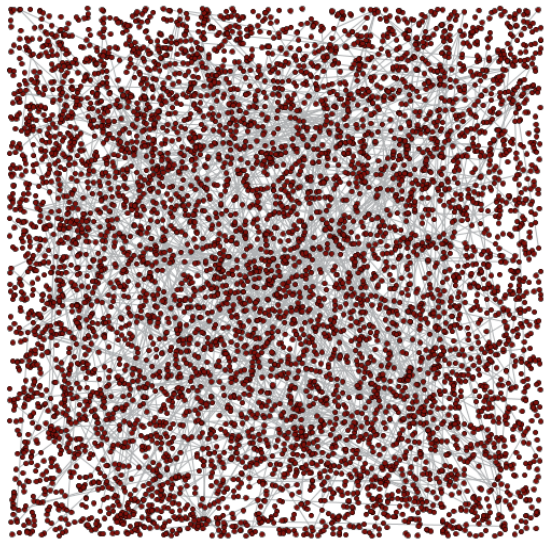

Figure 1 Erdos992 Pajek dataset (Batagelj and Mrvar 2006) drawn using Network WorkBench (Team 2006)

## 3. Cognitive Framework Development

In this section, we present the development of a formal agent-based modeling framework using a formal specification language (Spivey 1988).

**Towards a Formal**, Visual Framework of Emergent Cognitive Development of Scholars     9

The use of formal frameworks allows for exact specification of complex systems and logics as shown by Luck and Inverno for the domain of agents and multi-agent systems in (d'Inverno and Luck 2004). Another framework for development of agent-based models using a similar extension/description of formal specification for Wireless Sensor Networks sensing in Complex Adaptive Environments has been described in Niazi and Hussain (Niazi and Hussain 2010). The goal of the current framework is to describe and specify the agent-based model of researchers and its network representation using mathematical symbols instead of being a purely theoretical exercise as is common in formal specification models.

### 3.1 Framework

Before we can define a TCLN to represent state transitions over time, let us first define a network formally starting with the definition of a *Node*.

$$
\begin{array}{|l}
\hline
\textit{Node} \\
\hline
x, y: \mathbb{Z} \\
\hline
x \geq \textit{min-xcor} \\
x \leq \textit{max-xcor} \\
y \geq \textit{min-ycor} \\
y \leq \textit{max-ycor} \\
\hline
\end{array}
$$

So we have defined *Node* as containing x and y coordinates which are constrained by the limits of our Cartesian coordinate world. Next, we can define a schema for a *Line* as following

$$
\begin{array}{|l}
\hline
\textit{Line} \\
\hline
n1, n2: \textit{Node} \\
\hline
\end{array}
$$

A *Line* thus contains two nodes *n1* and *n2* of type *Node*.
Next, we can define a *Network* state schema as following:

$$
\begin{array}{|l}
\hline
\textit{Network} \\
\hline
\textit{nodes}: \mathbb{F}\,\textit{Node} \\
\textit{lines}: \mathbb{F}\,\textit{Line} \\
\hline
\forall\, x: \textit{lines} \,\exists\, y1, y2 \in \textit{nodes} \mid y1 = x.\,n1,\, y2 = x.n2 \\
\hline
\end{array}
$$

A network consists of lines and nodes, which we are depicting by including these two in the declaration. And in the predicate section, we are saying that for each line, there are two nodes, which are both members of the *nodes* subset. Next, we need to define a *ResearchPaper* schema as follows:

| *ResearchPaper* |
| --- |
| *Node* |
| *r:Researcher* |
| *citationCount*:$\mathbb{N}$ |
| *citationCount*$\geqslant$0 |

Now, here we are saying that *ResearchPaper* is going to be represented as a Node in our network. In addition, we are saying that we are including the Node and a variable of type *Researcher* schema in this schema. The problem here is that we have still not defined a *Researcher* till now.

So, we also need to formally specify the *Researcher* schema as following:

| *Researcher* |
| --- |
| *Node* |
| *rp*:$\{x|x\in ResearchPaper\}$ |
| *totalCites*:$\mathbb{N}$ |
| *h−index*:$\mathbb{N}$ |
| *totalCites*$\geqslant$0 |
| *x=h-index* |

Each *Researcher* firstly includes a *Node* and then finite set *rp* of *ResearchPaper* type. There is also a totalCites member of type "*fat N*", which represents the total number of citations that the *Researcher* has at a given time. Here, in the predicate section, we define the total citations to be a number greater than or equal to zero. A researcher also has an h-index value, which gives the Hirsch index at any given time. In addition, we want to place the *Researcher* on the X-Y plane based on the value of the Hirsch index, which we have accomplished by the schema inclusion of the *Node* schema.

Now that we have the basic framework, we can start with the definition of our specific novel "*TemporalCitationNetwork*" schema.

*TemporalCitationNetwork*
*Network*

$\forall x \in lines$

$x.y1 \Rightarrow x.y2 \in ResearchPaper$ and $x.y2 \in x.y1.rp$

or

$x.y2 \in Researcher \Rightarrow x.y1 \in ResearchPaper$ and $x.y1 \in x.y2.rp$

This specific type of *TemporalCitationNetwork* basically sets the rules that we have a network but for all lines in this network, the ends are a *ResearchPaper* and a *Researcher* only. This network is temporal because it is dependent upon time. The entire network evolves its state as *Researchers* publish more papers and papers get more citations. Let us define these state transitions as following:

*PublishPaper*

$r: \Delta Researcher$

$p?: Paper$

$\Delta TemporalCitationNetwork$

$r.rp' = r.rp \cup \{p?\}$

$lines' = lines \cup \{l\} \mid l.y1 = r, l.y2 = p?$

$r.TotalCites' = r.TotalCites + p?.CitationCount$

$r.h\_index' = UpdateHIndex(r)$

Here we can see that in the *PublishPaper* schema, we update the TCLN as well as the *Researcher* because of the addition of the new paper. Since this update occurs on an yearly basis, we can increase the citation count as well and subsequently use *UpdateHIndex*, an operation schema to update the Hirsch Index. A graphical representation of this can be seen in Figure 2.

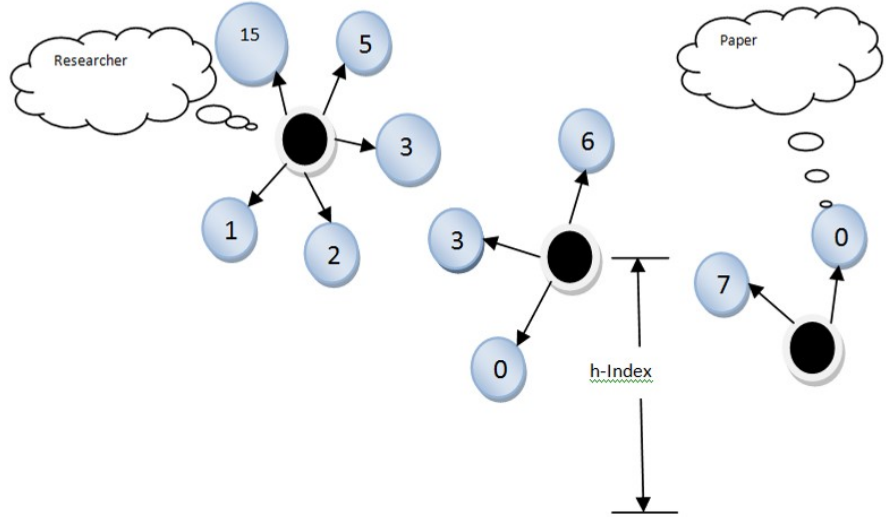

Figure 2 **TCLN with black nodes depicting researchers placed according to their Hirsch index and blue nodes depicting individual research papers and citations**

**Theorem 1: The use of TCLN requires space of the order of O(m) as compared to "traditional" author-paper Citation Networks whose space requirement is of the order of O(mn) ≈ O($n^2$), where "n" represents the number of average citations received per paper and "m" represents the average number of papers per researcher.**

**Proof**:
   The proof of the theorem is easy to follow. In case of traditional citation networks, each paper is connected to other papers. If we depict the papers and the papers citing them in the same network as in the case of traditional networks, then there will be inter-connections of the order of *mn*. On the other hand, in a TCLN, since we have reduced any such interconnections between researchers and other researchers or papers and other papers, each researcher and the published papers thus appear as a hub-spoke architecture as shown earlier in Figure 2. Thus the number of links to be drawn per researcher is equal to the number of papers. Hence this proves the theorem. The implications of the theorem 1 are also validated and can be seen in the agent-based simulation model described in the next section. □

## 4. Simulation Model

In this section, we present the agent-based simulation model for representation of TCLN. The model has been developed using NetLogo, a very popular agent-based simulation toolkit (Wilensky 1999). In NetLogo, an agent-based model typically consists of three different types of entities:
1. Turtles (or breeds of turtles),
2. Links (or breeds of links)
3. (optionally) interaction based on patches

Once these different entities have been defined, the key step in the model is to define the behavior of each agent. Agent-based modeling and simulation can be considered to be an advanced form of Object-Oriented programming. As such, it can be and is often used to represent complex concepts.

So, for our problem, as defined in the formal specification, we define the following types (*breeds*) of agents in our model:
A. Researcher agents
B. Research Paper agents

In the following sub-sections, we give an overview of these *breeds* of agents.

### 4.1 Researcher agents

These agents are made up of the breed "researchers". In NetLogo, this implies that we can subsequently address the breed separately. Now, similar to the model given in the formal specification, we enclose the following three attributes:

```
num-papers, total-citations and my-papers
```

As defined in the formal framework in Section 3, "num-papers" is the number of papers of the researcher. "my-papers" contains references to the ResearchPaper agents connected to this Researcher agent. "total-citations" represents an aggregation of all citations received by all papers. This is updated in the simulation on an "yearly" basic. Now, before we define the details of the interactions, let us also define the ResearchPaper agents.

### 4.2 ResearchPaper agents

These agents contain the following three variables

```
tend-to-be-cited ;tendency is from 0 to 1

num-cites ;Total number of citations

my-res; My author
```

"tend-to-be-cited" is a number representing the uniform probability of a paper to get citations. This is useful as an extension of the original model from the framework, because using this, the agent-based model can even represent growth of random researchers and not just real researchers based on retrieved data from indices. Being able to use randomly created set of researchers is done to allow for future extensions based on the testing of hypotheses such as those related to formation of research groups and journal editorial boards.

It would be pertinent to note here that in a TCLN, we shall show each paper as single-authored. In other words, each paper which has been co-authored will be reflected as many times as there are authors, in the model. The reason for choosing this is primarily to avoid any clutter and not show any primary structures of connectivity for optimization reasons (We refer to primary structure in a network to "direct inter-connections between nodes". An example of primary structure includes the original citation network format). The reason for this is that the goal of our work is not just to be able to represent the co-evolution of state-topology networks but also to reduce the memory space required to display the network visually (as proven earlier in Theorem 1). If this simplification is not performed, graphic rendering of an even small sub-section of a moderately complex citation network can range from a few minutes to hours, depending upon the available CPU and other hardware resources of the system. Thus, it would not be possible to observe the co-evolution of state-topology visually in real-time without the methodological simplification as proposed via TCLNs.

In a TCLN, with the passage of time, the key changes which can occur can be foreseen as following:

A.  The number of researcher agents increase over time
B.  The number of papers per researcher increases over time.
C.  The state of each paper (label) shows the citations and these increase over time,

What this really means is that the use of TCLN can greatly simplify the network representation and thus they are capable of display significantly more complex data easily.

### 4.3 Algorithms and functions

In this sub-section, we present the algorithms and functions implemented in the simulation for each agent. There are two parts of our agent-based model. One can be used to represent the evolution of an actual researcher by loading data from a file. The other part is relevant to generation of random researchers. Since the generation of the researchers is the more general case and the simulation of evolution of actual researchers is a specialized case, here we describe the general algorithms used.

Initially, in the *setup* function, we initialize the agents. Each researcher agent is initialized with a uniform random number based on the *max-init-papers* global variable controllable by a User Interface element called the "slider". Inside the same code, the agents then call *make-papers* and *set-lab* functions. Here, we show the code in the following segment in Figure 3. It is important to note here that we have purposefully chosen the actual code instead of other representations because of the complexity associated with agent-based algorithms. At times, we have observed that the native Logo code can be far more comprehensible as compared to traditional representations of code such as pseudo-code, flow-charts or sequence diagrams.

```
to make-papers

  hatch-papers num-papers
  [
    set shape "circle"
    set color blue
    set tend-to-be-cited ((random 101) + 1 ) / 100
    set num-cites random 10
    let temp [my-papers] of myself ;Add self to the list of the caller's paper
    set temp fput self temp
    set label num-cites
    ask myself
    [
      set my-papers temp
    ]
    create-link-with myself
    set heading random 360
    set my-res myself
    fd 3
  ]
end
```

**Figure 3** Code for the make-papers function

Here, we can see that each researcher agent calls "hatch" to create paper agents. These agents are then given a uniform random num-cites with initial citations less than ten. These are then each inserted in a list of papers retrieved from the researcher agent. Subsequently the list is updated again. Next, a link is created with the researcher agent and the paper agents align themselves spatially on the sides of the researcher.

Next, we show the algorithm for the calculation of h-index in Figure 4

```
to-report h-index
  let temp-list[]
  foreach my-papers
  [
    ask ?
    [
      set temp-list fput num-cites temp-list
    ]
  ]
  set temp-list reverse sort temp-list
  let i 0
  foreach temp-list
  [
    set i i + 1
    if i > ?
    [
      report i - 1
    ]
  ]
  if length temp-list = 1
  [
    report 1
  ]
  report i
end
```

Figure 4 **Code segment for the h-index algorithm**

As we can see, the function starts out by first sorting the papers according to their citations. Subsequently it performs a counting of the papers while also performing a comparison of the count with the current paper cites. If the index is greater than the paper then one less than that is reported as the Hirsch index of the researcher. And if no exact match is found, then after the loop is completed, then the count is reported as the Hirsch index. An exception is when the number of papers is just one, when the number "1" is simply the Hirsch index.

## 5. Validation Experiments

Validation is one of the most important steps in the development of any model. Validation of agent-based models (Niazi et al. 2010) can range from using collected data to ensure correlation with the real world to in-simulation validation such as using a Virtual Overlay Multi-agent System (Niazi et al. 2009). McBurney and White (McBurney and White 2006) note that validation is of four basic types:
1. Internal Validity
2. Construct Validity
3. External Validity
4. Statistical Validity

Internal validation refers to having sound reasons to believe that a cause-effect relation is present between independent and dependent variables. Statistical validity is similar to internal validity but also checks to see if the occurrence was not a pure chance. Construct validity refers to results supporting the theory behind the research and external validity refers to whether the results can be generalized to another situation.

Thus, according to this classification, we perform Internal, Construct and Statistical validation checks. However, we leave external validity to future work. Next, we show the validation of both the representational abilities of TCLNs as well as the calculation of the Hirsch index progression of researchers.

### 5.1 Validation of the representational abilities of TCLN models

To validate the random researcher generation capabilities of the agent-based model, we show the creation of n=60 random researchers in. Researchers are each placed on the Cartesian coordinate system, based on their Hirsch index. Each researcher, shown in black, has two labels. The first label shows the total number of published papers and the second shows the Hirsch index. Each paper, shown in blue, has the total citations as a label. This experiment validates the representational ability of the TCLN modeling paradigm for researchers as shown in Figure 5.

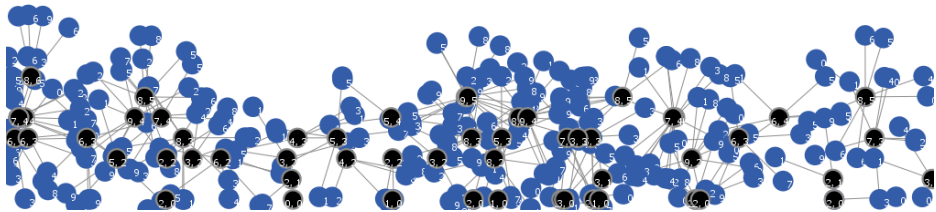

**Figure 5** Simulation view of n = 60 random researchers with max-init-papers = 10

### 5.2 Validation of Hirsch Index Calculation

For a further validation of TCLNs as well as the validation of the calculations of the Hirsch index, we take the emergent cognitive development of a renowned researcher "Victor Lesser", who is considered an authority in the domain of multi-agent systems. Using Publish or Perish ([Harzing and van der Wal 2009](#)), we queried Google scholar and discovered firstly that there are a total of 649 papers listed. However, 546 of these papers have actual years properly indexed so we shall use these in our analysis of the evolution of the researcher. The first paper shows itself in the year 1968 so we start there till 1988. The detailed h-index data for twenty years for "Victor Lesser" has been plotted in Figure 6.

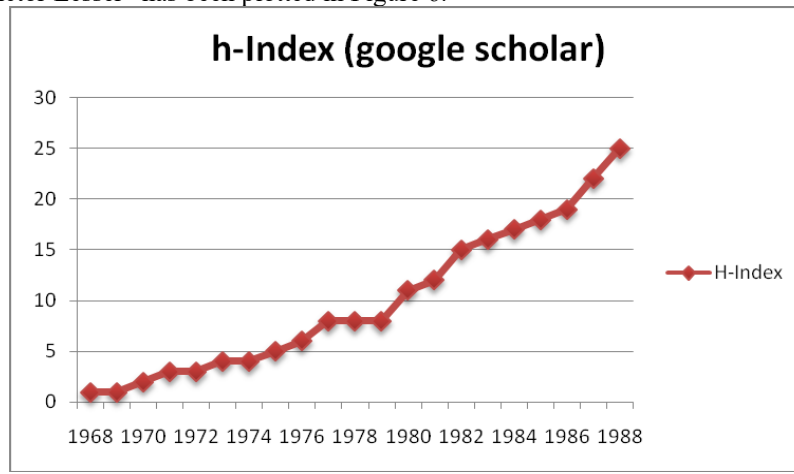

Figure 6 h-Index plot for twenty years for "Victor Lesser" obtained using Publish or Perish program

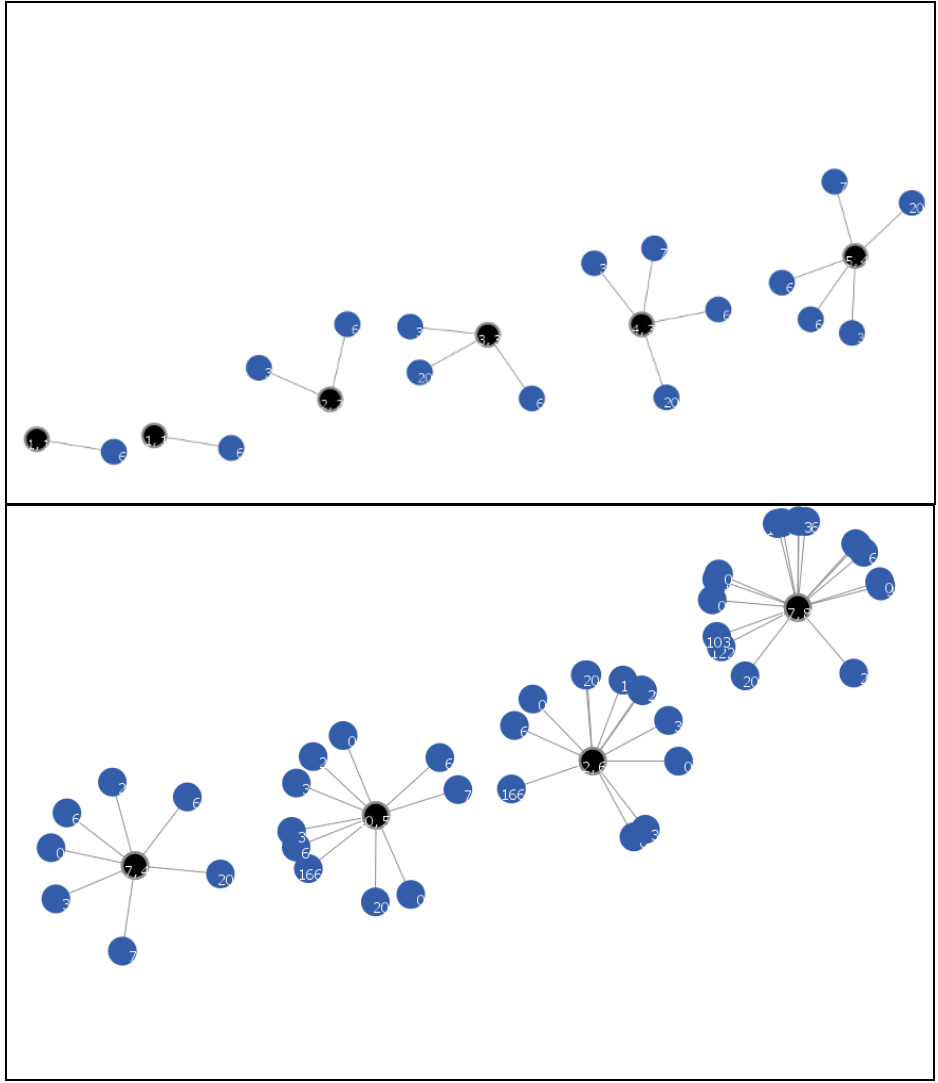

Figure 7 a, b Validation exercise 1: Evolution of a researcher's Hirsch-index (Victor Lesser)

Figure 7 a and b shows ten years of evolution of h-Index of Victor Lesser as depicted using TCLN using the simulated agent-based model. The detailed results and the retrieved data (via Google scholar index) is depicted in the experiments given in Table 1. In addition, we plot the number of nodes needed to be displayed in traditional citation networks versus TCLNs in Figure 8.

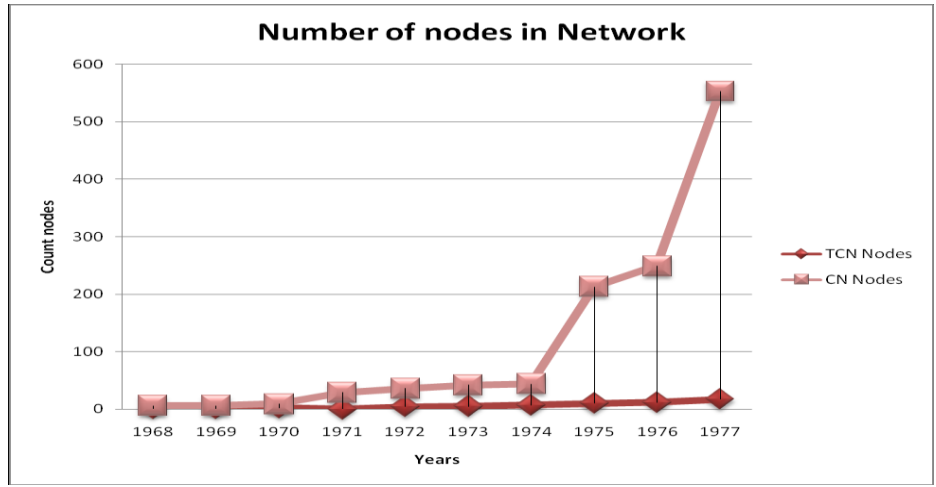

**Figure 8** Comparison of number of nodes needed to display a author-paper citation network vs. a TCLN for Victor Lesser's papers.

**Table 1 Data for validation. Calc hI = Calculated h-Index and h-Index is the h-Index from Publish or Perish software which obtains data via Google Scholar (Current as of 17 July, 2010)**

| Year | H-Index | Citations of papers | Calc hI |
|---|---|---|---|
| 1968 | 1 | 6 | 1 |
| 1969 | 1 | 6 | 1 |
| 1970 | 2 | [6 3] | 2 |
| 1971 | 3 | [20 6 3] | 3 |
| 1972 | 3 | [20 7 6 3] | 3 |
| 1973 | 4 | [20 7 6 6 3] | 4 |
| 1974 | 4 | [20 7 6 6 3 2 0] | 4 |
| 1975 | 5 | [166 20 7 6 6 3 3 2 0 0] | 5 |
| 1976 | 6 | [166 21 20 15 7 6 6 3 3 2 0 0] | 6 |
| 1977 | 8 | [166 122 103 71 21 20 15 8 7 6 6 3 3 2 0 0] | 8 |

Here, Table 1 shows two columns for the Hirsch index. One column shows the Hirsch index calculated using Google Scholar and the Publish or Perish program. On the other hand, the "Calc hI" column indicates the Hirsch index calculated by the algorithm alongside the evolution of the researcher. The table also shows the number of cited papers per researcher. Thus, we can see how we have validated both the calculation of the h-Index as well as the effectiveness of TCLNs for the modeling of researcher reputation (state) co-evolution with the changes in the topology using our agent-based model.

## 5.2 Discussion

In this section we first discuss the different validation techniques used in the framework. Next, we note how the proposed framework can be used for evaluating cognitive development of researchers across disciplines. We also discuss how the framework can be generalized for application in other domains.

### 5.2.1 Discussion of Validation

The first point to note here is the on statistical validation of any network model. Let us first examine how statistical validation is typically performed. In statistical validation, the data is sampled and compared with well-known data sources based on standard statistical metrics. However, this type of validation does not guarantee validation of entire results. Instead the results only match averages or standard deviation. Often statistical validation entails removing the outliers.

However, when we develop a complex network model, we basically take the entire dataset and develop the model around it. So, unlike statistical validation, this validation is completely truthful to the actual data since the network can simply be built only when every node and every link between the nodes has been developed expanded in the form of the network. In complex network models, each and every data point is further treated as a node/agent and thus is represented visually/formally in the form of a graph/network. Therefore any new measures for the measurement of any statistical data after the formation of this network would be an accurate and valid representation of the actual data.

Still, we would like to mention that the validity of any model can only be made as strong as the data it is validated against. So, for instance, if the citation indices such as Scholar/Web of Science etc. have errors inside them, then the same errors would be reflected in any model, whether statistical or network based. However, since in the network models, every node and link is of equal importance, such errors might show up with more clarity than models which aggregate the results and validate using statistical measures such as averages and curve-fitting.

### 5.2.2 Generalization of the Framework

The proposed framework can be considered as a means of representing the cognitive development of researchers. Even though, the framework has been demonstrated for the cognitive evolution of a Computer Science researcher, we have also shown in section 5.2 that the particular network representation is also useful for any set of researchers from any disciplines because the agent-based model based on NetLogo demonstrates the application of the modeling framework for any random researchers. In other words, the framework is actually a generalized application of computation techniques correlated with the Scientometric evolution of researchers aspiring from any discipline. Now, the interesting part here is that unless purely simulation-based models, the validation of the framework uses actual Scientometric data, available via Google Scholar, Scopus and Web of Science. As such, once we follow this particular modeling and simulation paradigm, it is trivial to model any researcher from any domain even in the same network. This would thus allow for the inter-disciplinary comparison of researchers using the proposed framework. Another possible generalization is by means of using a citation index similar to the individual h-index. Individual h-index (Batista et al. 2006) allows for measuring the notability of the author but reduces the effects of co-authorships.

### 5.2.3 Proposed Extensions of the Framework

In its current form, the framework only offers a view of representing the cognitive mental development of researchers from the point of view of other researchers. However, if we think about it, this could easily be considered as one application of the cognitive modeling paradigm. If we were to take another index such as g-index for the same scenario, we might come up with more interesting results. Unlike the h-index, the g-index gives more weight to highly cited articles. Likewise Schreiber's method (Schreiber 2008) of calculation of an index involves the use of fractional paper counts for multi-authored papers. The same modeling framework can be easily usable for any of these cases.

## 6. Conclusions and Future Work

In this paper, we have presented a formal modeling framework as well as the design, development and developed a prototype implementation of an agent-based simulation model for TCLN. We have shown how the use of our proposed TCLN model reduces the order of complexity as well as uses the Hirsch Index in the same network visualization model. We have validated our agent-based model by using a standard tool "Publish or Perish" to retrieve and calculate the evolution of h-Index over time for a high profile researcher starting from his first publication. We also validate the representational ability of TCLN by generating random researchers simulated by an agent-based model as well as simulation based on real data retrieved from the Google scholar citation index.

Although, the validation study uses the growth of a Computer Science researcher using Hirsch index, it is possible to use the same framework for researchers in other domains. In addition, TCLN representation is not limited to only the Hirsch index. Future research is planned for using the TCLN representation for modeling g-Index as well as hc-Index and others as proposed by the research community.

# Acknowledgement

We would like to express thanks to Dr. Tamim Khan at Bahria University for taking time to verify the formal framework developed in the paper.